\documentclass[prd,twocolumn,nofootinbib,aps,tightenlines,preprintnumbers,notitlepage,longbibliography,superscriptaddress]{revtex4-1}

\usepackage[dvipsnames]{xcolor}
\usepackage{color}
\usepackage{tikz}
\usepackage[caption=false]{subfig}
\usepackage{hyperref}
\usepackage{multirow}
\usepackage{siunitx}
\usepackage[export]{adjustbox}
\usepackage{graphicx}
\usepackage{dcolumn}
\usepackage{bm}
\usepackage[normalem]{ulem}
\usepackage{epstopdf}
\usepackage{tabularx}
\usepackage{suffix}
\usepackage{mathtools}
\usepackage{graphicx}
\usepackage{dcolumn}
\usepackage{bm}
\usepackage[normalem]{ulem}
\usepackage{xcolor}

\graphicspath{ {figures/} }

\usepackage{epstopdf}
\newcommand{\be}{\begin{eqnarray}}
\newcommand{\non}{\nonumber \\}
\newcommand{\ee}{\end{eqnarray}}

\newcommand{\nhat}{\hat{ \mathbf{n}}}

\def\spinup{\partial\kern-0.3em\raise0.42ex\hbox{\tiny\textbackslash}}
\def\spindown{\overline{\partial\kern-0.3em\raise0.42ex\hbox{\tiny\textbackslash}}}

\newcommand{\aap}{Astron. Astrophys.}

\newcommand{\jcap}{J. Cosm. Astropart. Phys.}

\newcommand{\dd}{{\rm d}}

\newcommand{\jhu}{William H. Miller III Department of Physics and Astronomy, 3400 N.\ Charles St., Baltimore, MD 21218, USA}

\begin{document}

\title{ Cosmological probes of helium reionization}

\author{Selim~C.~Hotinli}
\email{shotinl1@jhu.edu}
\affiliation{\jhu}

\date{\today}


\begin{abstract}

Joint analysis of CMB and large-scale structure at high redshifts provide new and unique windows into unexplored epochs of early structure formation. Here, we demonstrate how cosmic infrared background and high-redshift galaxies can be jointly analysed with CMB to probe the epoch of helium reionization ($2<z<4$) on the light cone using kinetic Sunyaev Zel'dovich tomography. Characterising this epoch has great potential significance for understanding astrophysics of galaxy formation, quasar activity and formation of the super-massive black holes. We find a detection at $8-10\sigma$ can be expected from combinations of data from CCAT-prime, Vera Rubin Observatory and CMB-S4 in the upcoming years.
    
\end{abstract}

\maketitle


\section{Introduction}

\vspace*{-0.25cm}

The epoch of large-scale ionization of the second electron in helium (hereafter `helium reionization') carries a large amount of information about astrophysics and cosmology. {Recently Ref.~\citep{Hotinli:2022jna} showed that the joint analysis of CMB and high-redshift ($2<z<4$) galaxy number-density fluctuations can be used to probe helium reionization via the technique of kinetic Sunyaev-Zel'dovich (kSZ) tomography (i.e. velocity reconstruction)~\citep[e.g.][]{Deutsch:2017ybc,Smith:2018bpn,Munchmeyer:2018eey,Zhang:2015uta,Hotinli:2019wdp,Cayuso:2019hen,Alvarez:2020gvl,Ferraro:2018izc, Smith:2016lnt, Hotinli:2020csk, AnilKumar:2022flx, Kumar:2022bly, Foreman:2022ves} and found a high-significance detection at  ${\sim10\sigma}$ can be expected from Vera Rubin Observatory (LSST)~\citep{LSST:2008ijt} and CMB-S4~\citep{CMB-S4:2016ple,CMB-S4:2020lpa} surveys in the near future. Here, we extend the calculations performed in Ref.~\citep{Hotinli:2022jna} to the more physically-motivated light-cone formalism and consider new and more complete set of large-scale structure (LSS) tracers, including the cosmic infrared background (CIB), high-redshift quasars and weak gravitational lensing of the CMB.}

{The detection of helium reionization proposed here relies on the increase in the electron fraction that results from ionizing helium. Since helium accounts for 8$\%$ of the baryonic nuclei by number (25$\%$ by mass), the number of electrons increases by an extra 8$\%$  compared to protons in the first reionization of helium (which occurs together with the hydrogen reionization), and then another 8$\%$  in the second reionization. Unlike reionization of hydrogen or the first reionization of helium, astrophysical models indicate photons emitted by the first stars are not sufficiently energetic to fully ionize the second electron in helium throughout the Universe.} The reionization of helium thus requires the emergence of large numbers of quasars or active galactic nuclei (AGN) therefore {occurs later and depends} strongly on the properties of quasars and AGN, such as their accretion mechanisms~\citep{Shen:2014rka}, luminosity function~\citep{Ross2013, Masters2012, McGreer2013, McGreer2018, Pan2022}, variability and lifetimes~\citep{Hopkins:2006vv, Schmidt2017}, as well as the growth and evolution of super-massive black holes~\citep{Inayoshi:2019fun}.

Probing helium reionization can also have implications for cosmology as the total change in the free-electron fraction during this time is a measure of the primordial helium abundance $Y_p$. Increasing the measurement accuracy on $Y_p$ may improve our understanding of the big bang nucleosynthesis, weak interaction rates, neutron lifetime (see e.g. for a review~~\citep{Pitrou:2018cgg}), as well as breaking the degeneracy between $Y_p$ and the number or relativistic degrees of freedom $N_{\rm eff}$~\citep{PhysRevD.87.083008}; potentially providing valuable insights into our cosmological history.

The precise details of helium reionization, such as its duration, timing and morphology are largely unknown. While surveys of helium and hydrogen Ly$\alpha$ forest can provide some evidence for the details of this epoch, these measurements--although likely more accurate in principle than the methods we discuss in this paper---are difficult in practice and are subject to severe astrophysical and systematic uncertainties about the inferred flux levels of the Ly$\alpha$ forest~\citep{2011MNRAS.410.1096B, 2014MNRAS.441.1916B, 2019ApJ...887..205T,Syphers2012}. Surveys of helium Ly$\alpha$, for example, are subject to intervening Lyman-limit systems at lower redshift \citep{Syphers2012}, significantly reducing the prospects of characterising the epoch of helium reionization unambiguously from these observables. Hydrogen Ly$\alpha$, on the other hand, provides only an indirect evidence for helium reionization through probing the thermal history of IGM, whose measurements are subject to significant systematic and modelling uncertainties~\citep{2011MNRAS.410.1096B, 2014MNRAS.441.1916B, 2019ApJ...887..205T}. These suggest additional probes of helium reionization we consider here will be valuable for increasing the prospects to unambiguously characterising this epoch. 

This paper is organised as follows: In Sec.~\ref{sec:cosmo_probes} we introduce various high-redshift  probes of large-scale structure such as the kSZ effect, CIB, weak-lensing of the CMB and high-redshift galaxies. In  Sec.~\ref{sec:velo_tomo} we describe the method of kSZ tomography using galaxies and CIB as tracers of the small-scale electron distribution. We describe our choices to  model upcoming CMB and LSS experiments and demonstrate the prospects of detecting velocities using  CIB in Sec.~\ref{sec:forecasts}. We assess the prospects to detect and characterise helium reionization from these probes using kSZ tomography in Sec.~\ref{sec:helium}. We conclude with a discussion of the path to probing helium reionization with joint analysis of CMB and LSS in Sec.~\ref{sec:discussion}.

\vspace*{-0.25cm}

\section{Cosmological probes of cosmic (fore)noon}\label{sec:cosmo_probes}

\vspace*{-0.25cm}

\subsection{The kSZ effect}

The temperature perturbation sourced due to the kSZ effect along the line of sight $\nhat$ takes the from
\be
{T(\nhat)|_{\rm kSZ}=-\sigma_T\int\dd\chi\,a\,n_e(\chi\nhat)v_r(\chi\nhat)}
\ee
where $\sigma_T$ is the Thomson scattering cross-section, $a$ is the scale factor, $\chi$ is the comoving distance, $n_e(\chi\nhat)$ is the {free} electron density along the line of sight, $v_r(\chi\nhat)$ is the remote dipole field at the electrons' rest frame, which we approximate to be equal to the peculiar radial velocity of electrons throughout this paper\footnote{In reality, Sachs-Wolfe (SW) effects also contribute to the remote dipole observed at the electrons' rest frame which can lead to a $\sim5-10$ percent effect on the angular power-spectrum of the remote dipole field on large scales and early redshifts~\citep{Deutsch:2017ybc,Cayuso:2018lhv,Hotinli:2020csk}.}. {The free electron density is proportional to $\bar{x}_e(z)$, the spatially-averaged free electron fraction,} which we define as the ratio of free electrons per hydrogen atom, which varies from zero to unity during reionization of hydrogen. The ionization of each electron in helium increases the ionization fraction by another $\sim8$ percent, set by the primordial helium abundance. 

\vspace*{-0.25cm}
\subsection{The CIB signal}\label{sec:CIB_signal}
\vspace*{-0.25cm}

The CIB signal is sourced by the thermal radiation of dust grains in distant star-forming galaxies. Dust grains absorb the ultraviolet starlight, heat up and re-emit light in the infrared. As star formation rate (SFR) of our Universe peak at around $z\sim2-3$, the CIB is sourced dominantly from galaxies at around these redshifts, coinciding with the epoch of helium reionization. Currently-available CIB maps provided by \textit{Planck}~\citep{aghanim2016planck} already allow CIB power-spectrum to be measured up to around sub-degree scales, sufficient for cosmological and astrophysical inference (see e.g. Refs~\citep{Mak:2016ykk,yu2017multitracer,Lenz:2019ugy,McCarthy:2022agq}), while upcoming measurements of CCAT-Prime will allow high-resolution measurements of CIB down to arc-minute scales~\citep{CCAT-Prime:2021lly}.

\begin{figure}[t!]
    \centering 
    \includegraphics[width = \linewidth]{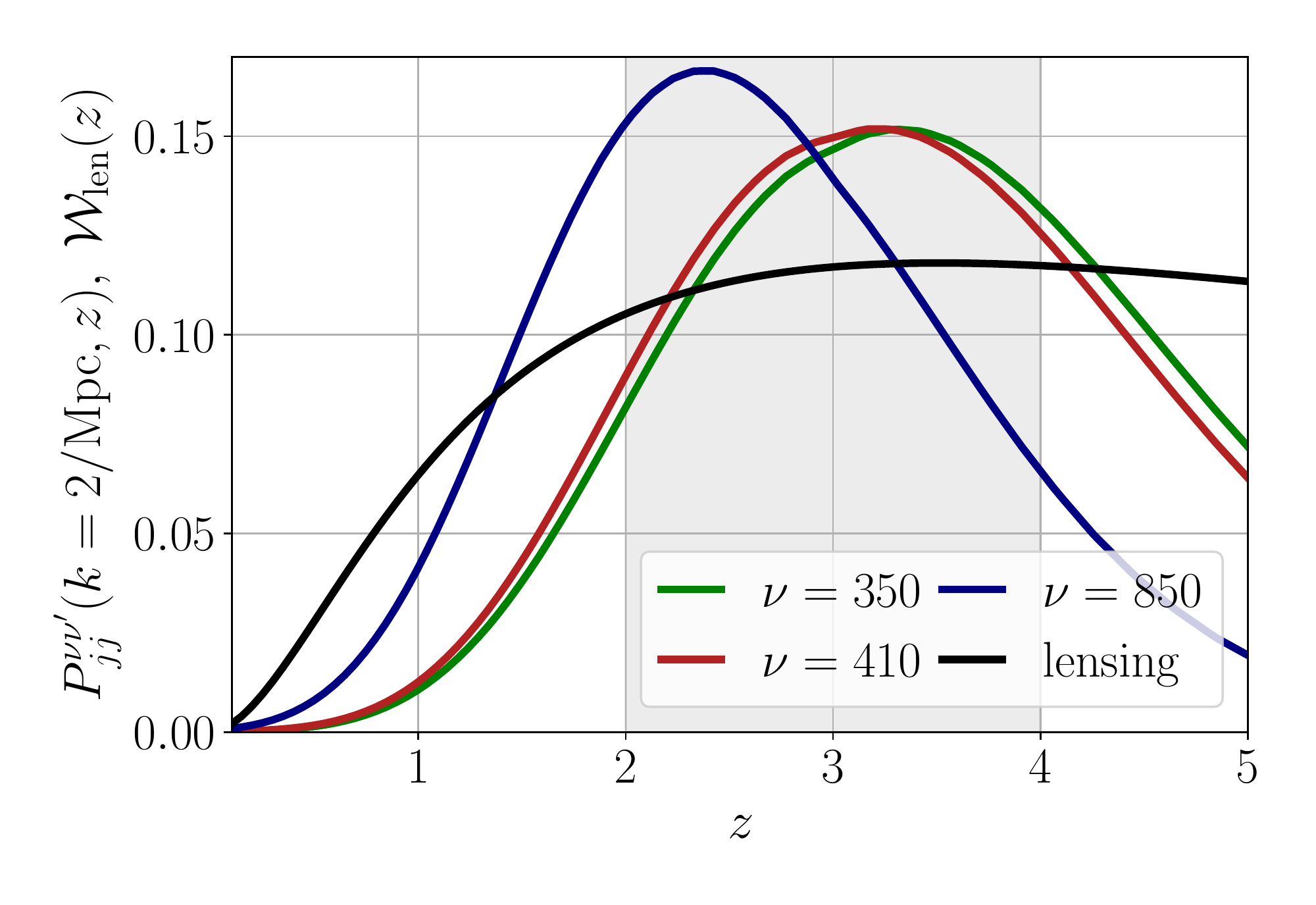}
    \vspace*{-1cm}
    \caption{Redshift dependence of the anticipated CIB brightness power-spectra $\nu^2\,(2\pi)^3\delta(\boldsymbol{k}\!-\!\boldsymbol{k}')P_{jj}^{\nu\nu'}(k,z)\!=\!\langle j_\nu(\boldsymbol{k})j_\nu(\boldsymbol{k}')\rangle$ at wavenumber $k=2
    {\rm Mpc}^{-1}$, shown together with the redshift kernel of weak gravitational lensing $
    \mathcal{W}_{\rm len}(z)\propto\chi(1-\chi/\chi_*)$. All curves are normalised  to equate to unity once integrated within  the redshift range $z\in[0.1,5.0]$. The light-gray--shaded region corresponds to the redshift range anticipated to correspond to the epoch of helium reionization, $2\lesssim z\lesssim 4$, where these signals can be seen to get significant contribution. Lines labelled as $\nu=\{350,410,850\}$ correspond to the anticipated observation frequencies of the upcoming CCAT-Prime survey.}
    \label{fig:CIBxCMB_signals-redshift}
    \vspace*{-0.5cm}
\end{figure}

The CIB brightness $I_\nu(\nhat)$ at frequency $\nu$ is given by the line-of-sight integral 
\be
I_\nu(\nhat)=\int_0^{\chi_*}\dd\chi a(\chi)j_\nu(\chi\nhat)\,,
\ee
where $j_\nu(\chi\nhat)$ is the emissivity density fluctuations which we write as $j_{\nu}(\chi\nhat)=\bar{j}_\nu(\chi)[1+\delta_{j_\nu}(\chi\nhat)]$ where $\delta_{j_\nu}(\chi\nhat)$ is the emissivity overdensity and $\bar{j}_\nu(\chi)$ is the mean emissivity density defined as an integral over the luminosity density 
\be
\bar{j}_{\nu_{\rm em}}(\chi)=\int\dd L_{\nu_{\rm em}}\frac{\dd N}{\dd L_{\nu_{\rm em}}}\frac{L_{\nu_{\rm em}}}{4\pi}\,,
\ee
where $\nu_{\rm em}=(1+z)\nu$ is the frequency corresponding to the redshift of the emitted radiation, $\dd N/\dd L_\nu$ is the luminosity function defined such that $\dd L_\nu (\dd N/\dd N_\nu)$ gives the number density of galaxies within luminosity between $L_\nu$ and $L_\nu+\dd L_\nu$. We model the CIB autospectra as
\be
C_{\ell}^{I_\nu I_\nu,\rm obs}=C_{\ell}^{I_\nu I_\nu}+N_\ell^{\nu,\rm SN}+N_\ell^{I_\nu I_\nu}\,,
\ee
where $C_{\ell}^{I_\nu I_\nu}$ is the CIB signal, $N_\ell^{\nu,\rm SN}$ is the shot noise term due to the finite number count of the galaxies sourcing the CIB signal and $N_\ell^{I_\nu I_\nu}$ is the instrumental noise of the CIB measurement to be defined in Sec.~\ref{sec:forecasts}. The CIB signal satisfy
\begin{equation}
\begin{split}
C_\ell^{I_\nu I_{\nu'}}&=\int\frac{2}{\pi}\int\dd\chi\dd\chi'\int k^2\dd k\\
&\times a(\chi)a(\chi')\bar{j}_\nu(\chi)\bar{j}_\nu(\chi')P_{jj}^{\nu\nu'}(k,\chi,\chi')j_{\ell}(k\chi)j_{\ell}(k\chi')\,,
\end{split}
\end{equation}
where $j_{\ell}(k\chi)$ is the spherical Bessel function and $(2\pi)^3\delta^3(\boldsymbol{k}\!-\!\boldsymbol{k}')P^{\nu\nu'}_{\delta_j\delta_j}(k,\chi,\chi')\!=\!\langle \delta_{j_\nu}\!(\boldsymbol{{k}},\chi') \delta_{j_{\nu'}}\!(\boldsymbol{k}',\chi')\rangle$ is the power-spectrum of the emissivity overdensity. 

We calculate the power-spectrum of the emissivity overdensity using the halo model, setting 
\be
P^{\nu\nu'}_{\delta_j\delta_j}(k,\chi,\chi')=P^{\nu\nu'\!,\rm 1h}_{\delta_j\delta_j}(k,\chi,\chi')+P^{\nu\nu'\!,\rm 2h}_{\delta_j\delta_j}(k,\chi,\chi')\,.
\ee 
Here, the 2-halo term $P^{\nu\nu'\!,\rm 2h}_{jj}(k,\chi,\chi')$ satisfy
\be
\bar{j}_\nu(\chi)\bar{j}_\nu(\chi')P^{\nu\nu',\rm 2h}_{\delta_j\delta_j}(k,\chi)=D_\nu(\chi)D_{\nu'}(\chi)P_{\rm lin}(k,z)\,,
\ee
where 
\be
D_\nu(\chi)\!=\!\frac{1}{{4\pi}}\int\dd M\frac{\dd N}{\dd M} b_{\rm h}(M,\chi)\left[{L_{v_{\rm em}}^{\rm cen}\!+\!L_{v_{\rm em}}^{\rm sat}u(k,M,\chi)}\right],\nonumber
\ee
is the CIB bias, $b_{\rm h}(M,\chi)$ is the halo bias, $\dd N/\dd M$ is the halo mass function, $P_{\rm lin}(k,z)$ is the linear matter-density power spectrum and $u(k, M, z)$ is the (normalised) Fourier transform of the halo density profile. Here $L_{v_{\rm em}}^{\rm cen}(M,z)$ and $L_{v_{\rm em}}^{\rm sat}(M,z)$ are the central and satellite galaxy luminosities respectively and we omit showing their mass and redshift dependencies in notation for brevity. The 1-halo term can be calculated as
\begin{equation}
\begin{split}
\bar{j}_\nu(\chi)\bar{j}_\nu(\chi')P^{\nu\nu',\rm 1h}_{\delta_j\delta_j}(k,\chi)=&\int\dd M\frac{\dd N}{\dd M}\frac{1}{(4\pi)^2}\\
&\times\left[L_{v_{\rm em}}^{\rm cen}L_{v_{\rm em}'}^{\rm sat} u(k,M,\chi)
\right.\\&\left.\ \ +L_{v_{\rm em}'}^{\rm cen}L_{v_{\rm em}}^{\rm sat} u(k,M,\chi)\right.
\\&\left.\ \ +L_{v_{\rm em}}^{\rm sat}L_{v_{\rm em}'}^{\rm sat} |u(k,M,\chi)|^2\right]\,,
\end{split}
\end{equation}
where the central-galaxy luminosity function $L_{v_{\rm em}}^{\rm cen}$ satisfy
\be
L_{v_{\rm em}}^{\rm cen}=N^{\rm cen}(M,z)L^{\rm gal}_{v_{\rm em}}(M,z)\,.
\ee
Here, $N^{\rm cen}(M,z)$ is the number of central galaxies in a halo of mass $M$ at redshift $z$, and $L^{\rm gal}_{v{\rm em}}(M,z)$ is the luminosity function of the host galaxy. The satellite galaxy luminosity function $L_{v_{\rm em}}^{\rm sat}$ is given by
\be
L_{v_{\rm em}}^{\rm sat}(M,z)=\int \dd M_s \frac{\dd N}{\dd M_s}L^{\rm gal}_{v_{\rm em}}(M_s,z)\,,
\ee
where $N^{\rm sat}=\int\dd M_s(\dd N/\dd M_s)$ and $\dd N/\dd M_s$ is subhalo mass function. Finally, the CIB shot noise can be written as
\be
N_{\ell}^{\nu,\rm NS}=\int\dd S_\nu\frac{\dd N}{\dd S_\nu}S_\nu^2\,,
\ee
where $S_\nu$ is the flux measured at frequency $\nu$. Our calculation of the CIB spectra follows Refs.~\citep{McCarthy:2019xwk,McCarthy:2020qjf} and the halo model described in Ref.~\citep{Smith:2018bpn}. {Throughout this paper we use the \texttt{ReCCO}\footnote{Publicly available at \href{https://github.com/jcayuso/ReCCO}{github/Recco}.} code (described in detail in Ref.~\citep{Cayuso:2021ljq}) when calculating observables.}

We show the CIB signal at various frequencies with blue solid curves on the left panel of Fig.~\ref{fig:signal_CIB_lensing}. There, the solid purple lines correspond to the total observed CIB signal including the shot noise and the detector noise. The dashed blue curves correspond to shot noise and the dotted purple curves correspond to detector noise. The solid gray and red lines correspond to CMB spectrum anticipated to be observed by Simons observatory and CMB-S4, respectively. We describe the experimental configurations used for noise contributions to these plots in Sec.~\ref{sec:forecasts}.

\subsection{Weak gravitational lensing of the CMB}
\vspace*{-0.25cm}

\begin{figure*}[t!]
    \centering
    \includegraphics[width = 0.661\linewidth]{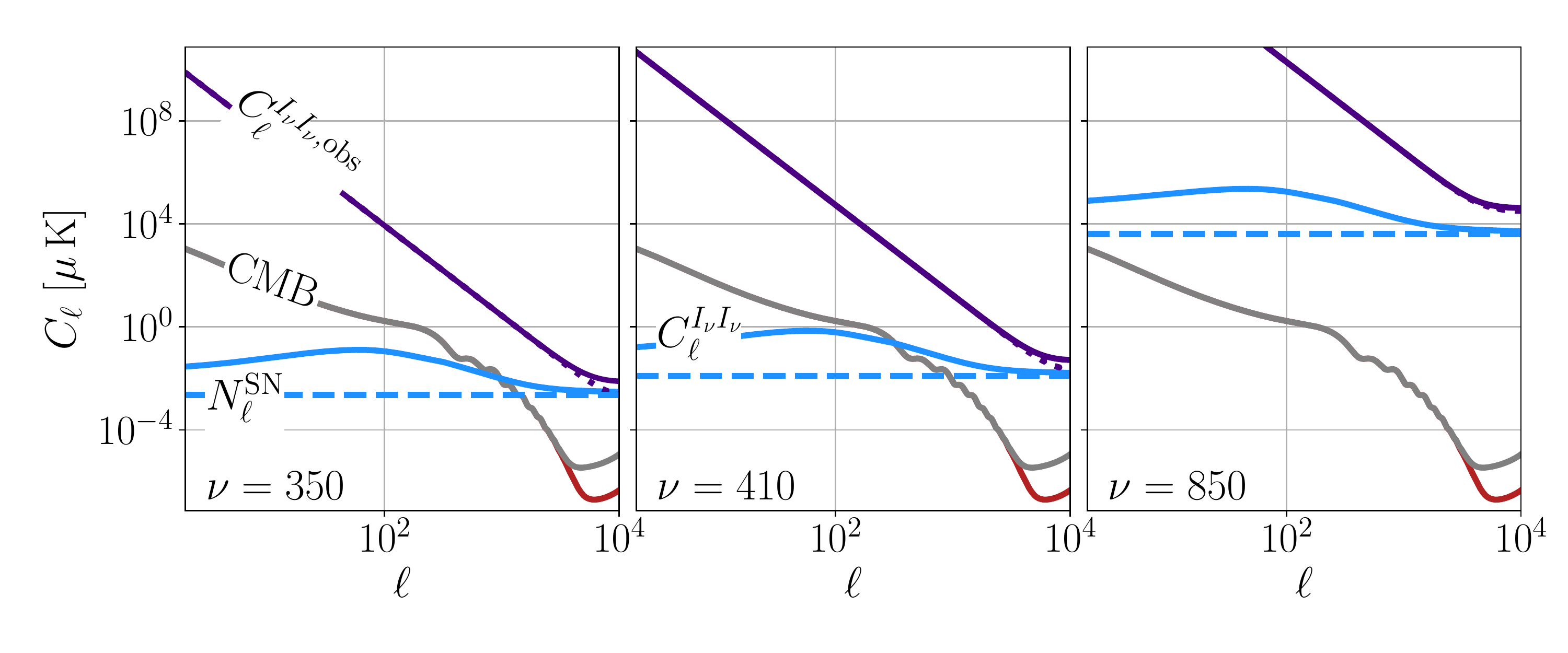}
    \hspace*{-0.1cm} \includegraphics[width = 0.333\linewidth]{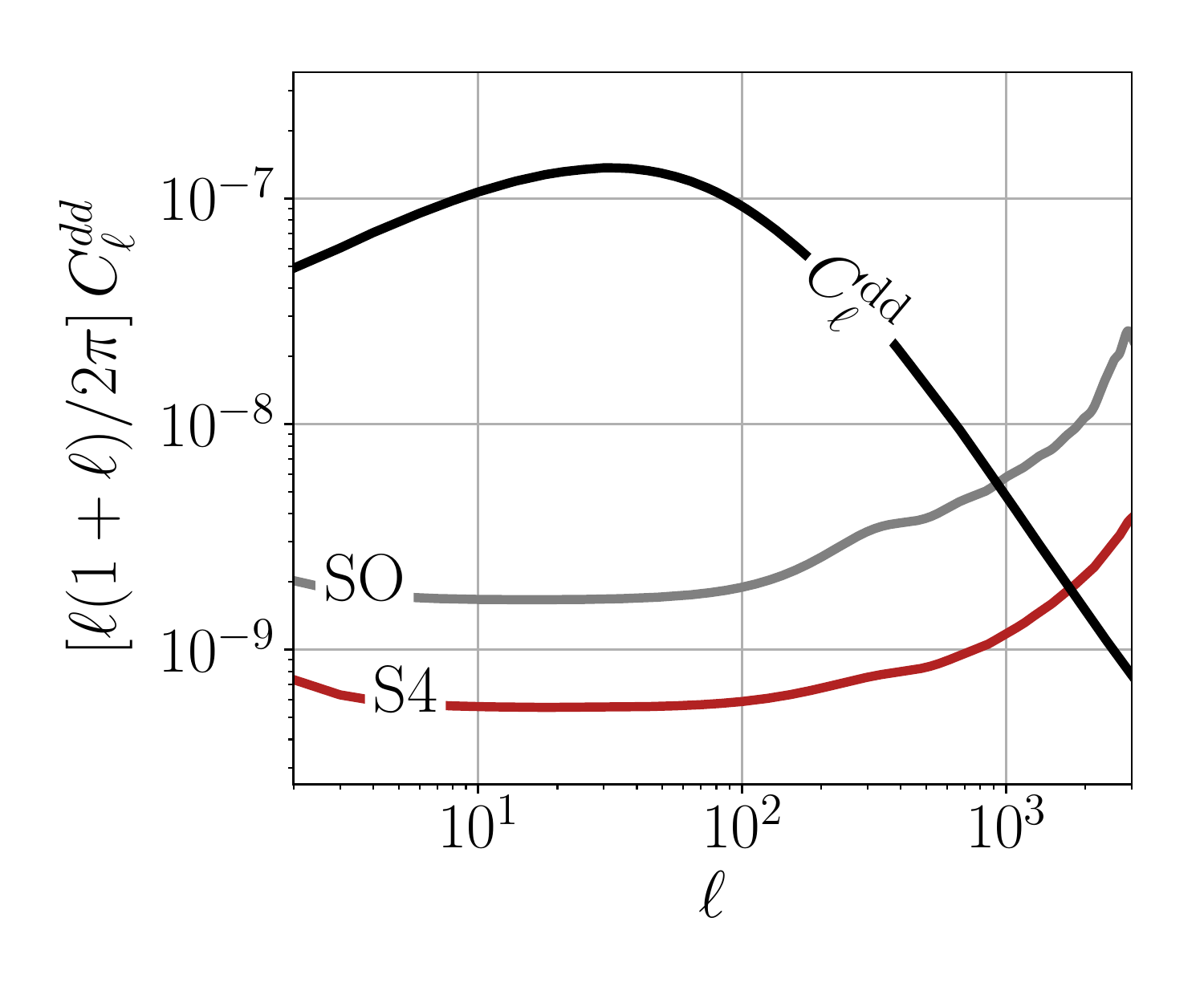}
    \vspace*{-1cm}
    \caption{(\textit{Left}) The anticipated signal and noise spectra as a function of multipoles $\ell$ for the CCAT-Prime measurements at frequencies $\nu=\{350,410,850\}$. The solid blue curves correspond to the anticipated CIB signal, the dashed blue curves correspond to the shot noise, the dotted purple lines correspond to the instrumental noise  and the {solid} purple lines correspond to the total spectra anticipated to be observed. {The instrumental noise is anticipated to dominate the observed spectra except at highest multiples.} The solid red and gray curves correspond to the  CMB signal including foregrounds and noise anticipated for CMB-S4 (S4) and Simons Observatory (SO) respectively. (\textit{Right}) The CMB weak-lensing deflection power-spectrum $C_\ell^{dd}$ shown together with the reconstruction-noise forecasts for CMB-S4 and Simons Observatory (SO).}
    \vspace*{-0.5cm}
    \label{fig:signal_CIB_lensing}
\end{figure*}

The CMB lensing potential is defined as
\be
\phi(\nhat)\equiv-2\int_0^{\chi_*}\dd\chi\frac{\chi^*-\chi}{\chi\chi^*}\Phi(\chi\nhat)\,,
\ee
where $\Phi(\chi\nhat)$ is the gravitational potential. The gravitational interaction of CMB photons and the large-scale structure intervening between the recombination surface and detectors on Earth deflects CMB photons by an angle given by $\boldsymbol{\alpha}=\boldsymbol{\nabla}\phi$. We show the anticipated lensing (deflection) power spectrum $C_\ell^{dd}=\ell(\ell+1)C_\ell^{\phi\phi}$ on the right panel of Fig.~\ref{fig:signal_CIB_lensing}, along with the lensing-reconstruction noise calculated using the standard lensing quadratic estimator following~\citep{Hotinli:2021umk}. Here, we used the \texttt{class\_delens}\footnote{Publicly available at \href{https://github.com/selimhotinli/class_delens}{github/selimhotinli/class\_delens}.} code to calculate the lensing reconstruction noise, assuming experimental configurations matching Simons Observatory and CMB-S4, which are described below. The redshift dependence of the lensing signal $\mathcal{W}_{\rm len}(z)\propto \chi(1-\chi/\chi^*)$ was shown in Fig.~\ref{fig:CIBxCMB_signals-redshift}, suggesting lensing gets significant contribution from high redshifts corresponding to the epoch of helium reionization. The cross power of lensing deflection with the radial velocity and galaxy fluctuations are given in Ref.~\citep{Cayuso:2021ljq}.

\vspace*{-0.25cm}
\subsection{Galaxies}
\vspace*{-0.25cm}

Similar to the bin-averaged radial-velocity and optical-depth fields, we construct a 2-dimensional galaxy density field as an integral over a given redshift bin as
\be
\delta_g^\alpha(\nhat)=\int_{\chi_\alpha^{\rm min}}^{\chi_\alpha^{\rm max}}\dd\chi_\alpha \mathcal{W}_{\rm gal}(\chi_\alpha)\delta_g(\chi_\alpha\nhat)\,,
\ee
where 
\be
\mathcal{W}_{\rm gal}({\chi_\alpha})=H(z_\alpha)\int\dd\chi W(\chi)P(z_\alpha,z(\chi))\,,
\ee
is the effective window function taking into account the photo-$z$ errors, and
\be
P(z,z_\alpha)=\bar{P}(z)\exp\left[-\frac{(z_\alpha-z)^2}{2\sigma_z}\right]\,,
\ee
with $\bar{P}(z)=\int_0^{\infty}\dd z'\exp[-{(z'-z)^2}/{2\sigma_z}]$. Here, $\sigma_z$ is the anticipated photo-$z$ error for a given galaxy survey.

The observed angular power
spectrum between redshift-binned galaxy density fluctuations from a photometric survey can then be expressed as
\begin{equation}
\begin{split}
&C_{\ell,\alpha\beta}^{\delta_g\delta_g, \rm obs}\!\!\!=\!\!16\pi^2\!\!\int_{\chi_\alpha^{\rm min}}^{\chi_\alpha^{\rm max}}\!\!\!\!\!\!\dd\chi_\alpha\int_{\chi_\beta^{\rm min}}^{\chi_\beta^{\rm max}}\!\!\!\!\!\!\dd\chi_\beta \,\,\mathcal{W}_{\rm gal}(\chi_\alpha)\mathcal{W}_{\rm gal}(\chi_\beta) \\ 
&\times\!\!\!\int\!\frac{\dd k k^2}{(2\pi)^3}j_\ell(k\chi_\alpha)j_\ell(k\chi_\beta) P_{gg}(\chi_\alpha,\chi_\beta,k)\!+\!\delta_{\alpha\beta}\frac{1}{n_g(z_\alpha)}\,,
\end{split}
\end{equation}
where $P_{gg}(\chi_\alpha,\chi_\beta,k)$ is the galaxy-galaxy power spectrum which we compute using the halo model as prescribed in Ref.~\citep{Cayuso:2021ljq}. Here, $n_{\rm gal}(z_\alpha)$ is the galaxy shot noise which we define in Sec.~\ref{sec:forecasts}. The cross power of the galaxy and radial velocity fluctuations on large scales is defined in Ref.~\citep{Cayuso:2021ljq}.

\section{Velocity reconstruction}\label{sec:velo_tomo}

\subsection{kSZ tomography}

An important consequence of the kSZ effect is that the cross power of the CMB and a tracer of electron fluctuations (such as the distributions of galaxies or the CIB signal) becomes anisotropic on small-scales; varying over the sky in a way dependent on the fluctuations of the bulk radial velocities of electrons. The cross correlation of the kSZ effect and a tracer of density fluctuations such as the galaxy overdensity $\delta_g^\beta(\chi\nhat)$ inside some redshift bin $\beta$ can be written on the curved sky in terms of spherical harmonic coefficients as
\be\label{eq:cross_CIB_1}
\langle T_{\ell m}\delta^\beta_{\ell' m'}\rangle&=&\sum\limits_\alpha\sum\limits_{LML'M'}\bar{v}^\alpha_{L'M'}\langle\tau^\alpha_{LM}\delta^\beta_{g,L'M'}\rangle\\ &\times&\int\dd^2\nhat Y_{\ell m}^*(\nhat)Y_{LM}(\nhat)Y_{L'M'}(\nhat)\,,\nonumber
\ee
where $\langle \tau^\alpha_{\ell m} \delta^{*\beta}_{\ell'm'}\rangle\equiv C_\ell^{\tau_\alpha \delta_\beta}\delta_{\ell\ell'}\delta_{mm'}$. Here $\bar{v}^\alpha_{LM}$ is the spherical-harmonic transform of the mean radial velocity field averaged over a redshift bin. The radial-velocity field at the comoving distance $\chi_\alpha$ can also be parametrised as $v_r(\chi\nhat)=\bar{v}_r^\alpha(\nhat)[1+\delta v_r (\chi_\alpha\nhat)]$ where $\bar{v}_r^\alpha$ radial-velocity field averaged within the comoving distance range $[\chi_\alpha^{\rm min},\chi_\alpha^{\rm max}]$ and satisfy
\be
\bar{v}_r^\alpha(\nhat)=\frac{1}{\Delta\chi_\alpha}\int_{\chi_\alpha^{\rm min}}^{\chi_\alpha^{\rm max}}\dd\chi_\alpha v_r(\chi_\alpha\nhat)\,,
\ee
and $\bar{v}^{\alpha}_{LM}\equiv\int\dd\nhat\,\bar{v}_r^{\alpha}(\nhat)Y_{LM}(\nhat)$. We can similarly write the electron density field as $n_e(\chi\nhat)=\bar{n}_e(\chi)[1+\delta_e(\chi\nhat)]$ where $\bar{n}_e(\chi)$ is the sky-averaged electron density at the comoving distance $\chi$ and $\delta_e(\chi\nhat)$ is the fluctuations of electron overdensity.  The anisotropies in the bin-averaged optical depth then satisfy 
\be
\tau^\alpha(\nhat)=-\sigma_T\int_{\chi_\alpha^{\rm min}}^{\chi_\alpha^{\rm max}}\dd\chi\,a\,\bar{n}_e(\chi)[1+\delta_e(\chi\nhat)]\,,
\ee
where $\tau_{LM}^{\alpha}\equiv\int\dd\,\nhat \tau_r^{\alpha}(\nhat)Y_{LM}(\nhat)$ in Eq.~\eqref{eq:cross_CIB}, and the mean-field contributions to the kSZ signal in the CMB take the form
\be
T(\nhat)|_{\rm kSZ}=\sum\limits_\alpha \tau^\alpha(\nhat)\bar{v}_r^\alpha(\nhat)\,.
\ee
Rewriting the second line of Eq.~\eqref{eq:cross_CIB_1} using Wigner-3J symbols we get
\be
\langle T_{\ell m}\delta^\beta_{\ell'm'}\rangle\!=\!\!\!\!\sum\limits_{\alpha,L'M'}\!\!\!(-1)^{m_1+m_2}\Gamma^{\alpha\beta}_{\ell\ell'L'}
\begin{pmatrix}
\ell & \ell' & L' \\
 \!m & \!-m' & M'
\end{pmatrix}\bar{v}^{\alpha}_{LM}\,\,\,
\ee
where 
\be
\Gamma^{\alpha\beta}_{\ell\ell'L'}\!\equiv\!\sqrt{\frac{(2\ell+1)(2\ell'+1)(2L'+1)}{4\pi}}\!\begin{pmatrix}
\ell & \ell' & L' \\
 \!0 & \!0 & 0
\end{pmatrix}C_{L'}^{\tau^\alpha \delta_g^\beta}\!\!.\,\,\,\,
\ee

The program of reconstructing the large-scale radial-velocity field from this statistical anisotropy is called `kSZ tomography' where the unbiased and minimum-variance quadratic estimator for the redshift-binned bulk velocity field takes the form
\be
\hat{\bar{v}}_{\ell m}^{\alpha}=A_\ell^{\alpha}(-1)^m\!\!\!\!\sum\limits_{\ell'm'LM}\!\!\!\begin{pmatrix}
\ell & \ell' & L' \\
 \!0 & \!0 & 0
\end{pmatrix}\Gamma^{\alpha\beta}_{\ell'L\ell}\frac{T_{\ell'm'}\delta^\beta_{L M}}{C_{\ell'}^{TT}C_{L}^{\delta_g^\beta\delta_g^\beta}}\,,
\ee
with the reconstruction noise of the estimator satisfying
\be\label{eq:noise_kSZ_gal}
N_{\alpha\ell}^{\bar{v}\bar{v},\rm kSZ}=\left[\frac{1}{2\ell+1}\sum\limits_{\ell'L}\frac{\Gamma^{\alpha\beta}_{\ell'L\ell}\Gamma^{\alpha\beta}_{\ell'L\ell}}{C_{\ell'}^{TT,\rm obs}C_L^{\delta_g^\beta\delta_g^\beta,\rm obs}}\right]^{-1}.
\ee
The nominator of the summed term inside the brackets in Eq.~\eqref{eq:noise_kSZ_gal} contains a product of two Wigner-3J symbols which can be written as an integral of the product of three Wigner-d matrices using the equality
\begin{equation}\label{eq:spherical_dmatrix}
\begin{split}
\int_{-1}^{1}\dd(\cos{\theta})\ d^{\ell_1}_{s_1s'_1}&(\theta)d^{\ell_2}_{s_2s'_2}\!(\theta)d^{\ell_3}_{s_3s'_3}\!(\theta)\\ &=2\begin{pmatrix}
    \ell_1 & \ell_2 & \ell_3 \\ 
    s_1 & s_2 & s_3 \\ 
    \end{pmatrix} 
    \begin{pmatrix} 
    \ell_1 & \ell_2 & \ell_3 \\ 
    s'_1 & s'_2 & s'_3 \\ 
    \end{pmatrix}\,,
\end{split}
\end{equation}
which gives
\be
N_{\alpha\ell}^{\bar{v}\bar{v},\rm kSZ}=2\pi \int_{-1}^{1}\!\!\!\dd(\cos{\theta})\zeta_1(\theta)\zeta^{\alpha\beta}_2(\theta)d_{00}^{\ell}(\theta)\,,
\ee
where
\be
\zeta_1(\theta)=\sum\limits_{\ell}\frac{(2\ell+1)}{4\pi}\frac{1}{C_{\ell}^{TT,\rm obs}}\,,
\ee
and
\be
\zeta_2^{\alpha\beta}(\theta)=\sum\limits_{\ell}\frac{(2\ell+1)}{4\pi}\frac{(C_{\ell,\alpha\beta}^{\tau\delta_g})^2}{C_{\ell,\beta\beta}^{\delta_g\delta_g,\rm obs}}\,.
\ee
Here, the cross-correlation between the redshift-bin averaged optical depth and galaxy fields satisfy
\begin{equation}
\begin{split}
&C_{\ell,\alpha\beta}^{\tau\delta_g}\!=\!\! 16\pi^2\sigma_T\!\!\int_{\chi_\alpha^{\rm min}}^{\chi_\alpha^{\rm max}}\!\!\!\!\!\!\dd\chi_\alpha\!\!\int_{\chi_\beta^{\rm min}}^{\chi_\beta^{\rm max}}\!\!\!\!\!\!\dd\chi_\beta \,\,a(\chi_\alpha)\bar{n}_e(\chi_\alpha) \mathcal{W}_{\rm gal}(\chi_\beta)\\ 
&\,\,\,\,\,\,\,\,\,\,\,\,\times\!
\!\!\int\!\frac{\dd k k^2}{(2\pi)^3}j_\ell(k\chi_\alpha)j_\ell(k\chi_\beta)P_{eg}(\chi_\alpha,\chi_\beta,k)\,,
\end{split}
\end{equation}
where $P_{eg}(\chi_\alpha,\chi_\beta,k)$ is the electron-galaxy cross power, determined by the model for the electron density profile inside dark matter halos, which depends on the physical
processes such as AGN feedback. Throughout this paper we choose the `AGN' gas profile model from Ref.~\citep{Battaglia:2016xbi} to model electron profiles and use the \texttt{ReCCO} code for our calculations following Ref.~\citep{Cayuso:2021ljq}.

\subsection{Velocity reconstruction with the CIB}

The cross-correlation of the kSZ effect and CIB intensity can be written as 
\be\label{eq:cross_CIB}
\langle T_{\ell m}I_{\nu,\ell' m'}\rangle&=&\sum\limits_\alpha\!\!\!\sum\limits_{LML'M'}\!\!\bar{v}^\alpha_{L'M'}\langle\tau^\alpha_{LM}I_{\nu,L'M'}\rangle\\ &\times&\int\dd^2\nhat Y_{\ell m}^*(\nhat)Y_{LM}(\nhat)Y_{L'M'}(\nhat)\nonumber
\ee
where $\langle \tau^\alpha_{\ell m} I^{*}_{\nu,\ell'm'}\rangle\equiv C_\ell^{\tau_\alpha I_\nu}\delta_{\ell\ell'}\delta_{mm'}$. Rewriting the second line of Eq.~\eqref{eq:cross_CIB} with Wigner-3J symbols we get
\be
\langle T_{\ell m}I_{\ell'm'}\rangle=\!\!\!\!\sum\limits_{\alpha L'M'}\!\!(-1)^{m_1+m_2}\Gamma^{\alpha,\rm CIB}_{\ell\ell'L'}
\begin{pmatrix}
\ell & \ell' & L' \\
 \!m & \!-m' & M'
\end{pmatrix}\bar{v}^{\alpha}_{LM}\,,\non
\ee
where 
\be
\Gamma^{\alpha,\rm CIB}_{\ell\ell'L'}\!\equiv\!\sqrt{\frac{(2\ell+1)(2\ell'+1)(2L'+1)}{4\pi}}\!\begin{pmatrix}
\ell & \ell' & L' \\
 \!0 & \!0 & 0
\end{pmatrix}C_{L',\alpha}^{\tau I_\nu}.\,\,\non
\ee
Here, the cross-correlation between the bin-averaged optical depth and the CIB brightness can be written as 
\begin{equation}
\begin{split}
&C_{\ell,\alpha}^{\tau I_\nu}= 16\pi^2\sigma_T\int_{\chi_\alpha^{\rm min}}^{\chi_\alpha^{\rm max}}\!\!\!\!\!\!\dd\chi_\alpha\int_{0}^{\chi_*}\!\!\!\!\!\!\dd\chi_\beta \,\,j_\ell(k\chi_\alpha)j_\ell(k\chi_\beta)\\ 
&\,\,\,\,\,\,\,\,\,\,\,\,\times\!a(\chi_\alpha)\bar{n}_e(\chi_\alpha) a(\chi_\beta)\bar{j}_\nu(\chi_\beta)\!\!\int\!\frac{\dd k k^2}{(2\pi)^3} P_{ej}^\nu (\chi_\alpha,\chi_\beta,k)
\end{split}
\end{equation}
where $(2\pi)^3\delta^3(\boldsymbol{k}\!-\!\boldsymbol{k}')P^{\nu}_{ej}(k,\chi,\chi')\!=\!\langle \delta_{e}(\boldsymbol{{k}},\chi') \delta_{j_{\nu}}(\boldsymbol{k}',\chi')\rangle$ is the cross power between the fluctuations in CIB emissivity and electron density. Following Ref.~\citep{Cayuso:2021ljq}, we calculate the 2-halo contribution to this cross power as 
\be
P^{\nu,{\rm 2h}}_{XY}(k,z)=D_X(k,z)D_Y(k,z)P_{\rm lin}(k,z)\,,
\ee
where $P_{\rm lin}(k,z)$ is the linear dark matter power spectrum and 
\be
D_X(k,z)=\int\dd M\frac{\dd N}{\dd M} b_h(M,z)A_X\,.
\ee
For the electron and emissivity fluctuations, these satisfy
\be
A_e=\frac{M}{\rho_m} u_e(k,M,z)\,,
\ee
and
\be
A_j=\frac{1}{4\pi}\left[{L_{v_{\rm em}}^{\rm cen}\!+\!L_{v_{\rm em}}^{\rm sat}u(k,M,\chi)}\right]\,,
\ee
where $u_e(k,M,z)$ is the electron density profile, which we set to the `AGN' gas profile defined in Ref.~\citep{Battaglia2016}, and $\rho_m$ is the present day cosmological matter density. The $1$-halo term for the cross-correlation can be calculated as
\be
P^{\nu,{\rm 1h}}_{ej}(k,z)=\int \dd M \frac{\dd N}{\dd M} A_e(M,k,z) A_j(M,k,z)\,.
\ee

Similar to the case of kSZ tomography with galaxies, we can now define a minimum-variance quadratic estimator for the bulk velocity field at some redshift bin $\alpha$. Since both CMB and CIB are 2-dimensional fields integrated along the line of sight, we first drop the constraint that the reconstructed redshift-binned velocity field must be unbiased. The \textit{biased} minimum variance estimator can be written as 
\be
\hat{\bar{v}}_{\ell m}^{(b)\alpha}=A_\ell^{\alpha}(-1)^m\sum\limits_{\ell'm'LM}\begin{pmatrix}
\ell & \ell' & L' \\
 \!0 & \!0 & 0
\end{pmatrix}\Gamma^{\alpha,\rm CIB}_{\ell'L\ell}\frac{T_{\ell'm'}I_{\nu,L M}}{C_{\ell'}^{TT}C_{L}^{I_\nu I_\nu}}\,,\non
\ee
where
\be
A^{\alpha}_\ell=\left[\frac{1}{2\ell+1}\sum\limits_{\ell'L}\frac{\Gamma^{\alpha,\rm CIB}_{\ell'L\ell}\Gamma^{\alpha,\rm CIB}_{\ell'L\ell}}{C_{\ell'}^{TT}C_{L}^{I_\nu I_\nu}}\right]^{-1}\,.
\ee
The velocity reconstruction noise then satisfies
\be
N_{\ell,\alpha\beta}^{(b)\bar{v}\bar{v}}=\frac{A^\alpha_\ell A_\ell^\beta}{2\ell+1}\sum\limits_{\ell'L} \frac{\Gamma^{\alpha,\rm CIB}_{\ell'L\ell} \Gamma^{\beta,\rm CIB}_{\ell'L\ell}}{C_{\ell'}^{TT}C_{L}^{I_\nu I_\nu}}\,.
\ee

We can now define an \textit{unbiased} quadratic estimator for the velocity as 
\be
\hat{\bar{v}}_{\ell m}^{\alpha}=(R^{-1})_{\alpha\beta}\hat{v}_{\ell m}^\beta\,,
\ee
which satisfies $\langle \hat{v}_{\ell m}^{\alpha}\rangle=v_{\ell m}^{\alpha}$. The rotation matrix that de-biases the reconstructed velocity can be found to satisfy
\be
R_{\alpha\beta}=\left[\sum\limits_{\ell'L}\frac{\Gamma^{\alpha,\rm CIB}_{\ell'L\ell}\Gamma^{\alpha,\rm CIB}_{\ell'L\ell}}{C_{\ell'}^{TT}C_{L}^{I_\nu I_\nu}}\right]^{-1}\sum\limits_{\ell'L}\frac{\Gamma^{\alpha,\rm CIB}_{\ell'L\ell}\Gamma^{\beta,\rm CIB}_{\ell'L\ell}}{C_{\ell'}^{TT}C_{L}^{I_\nu I_\nu}}\,.
\ee
The reconstruction noise for the unbiased minimum variance estimator satisfy
\be\label{eq:CIB_kSZ_unbiased}
N_{\ell,\alpha\beta}^{\bar{v}\bar{v}}=(R^{-1})_{\alpha\gamma}(R^{-1})_{\beta\delta}N_{\ell,\gamma\delta}^{(b)\bar{v}\bar{v}}\delta_{\ell\ell'}\delta_{mm'}.
\ee

Finally, using the spherical-harmonic equality defined in Eq.~\eqref{eq:spherical_dmatrix}, we rewrite the (biased) reconstruction noise as
\be\label{eq:noise_rec_CIB}
N_{\ell,\alpha\beta}^{(b)\bar{v}\bar{v}}=2\pi\frac{A^\alpha_\ell A_\ell^\beta}{2\ell+1}\int_{-1}^{1}\!\!\!\dd(\cos{\theta})\zeta_1(\theta)\zeta^{\alpha\beta}_{2,\rm CIB}(\theta)d_{00}^{\ell}(\theta)\,,
\ee
where 
\be
\zeta^{\alpha\beta}_{2,\rm CIB}(\theta)=\sum\limits_\ell\frac{(2\ell+1)}{4\pi}\frac{C_{\ell,\alpha}^{\tau I_\nu}C_{\ell,\beta}^{\tau I_\nu}}{C_{\ell}^{I_\nu I_\nu,\rm obs}}\,,
\ee
and
\be
A_\ell^\alpha=2\pi\int_{-1}^{1}\!\!\!\dd(\cos{\theta})\zeta_1(\theta)\zeta^{\alpha\alpha}_{2,\rm CIB}(\theta)d_{00}^{\ell}(\theta)\,.
\ee

\section{Forecasts}\label{sec:forecasts}

Throughout this paper we use the standard \textit{Planck} cosmology with 6 $\Lambda$CDM model parameters we define in Table~\ref{table:cosmo_fiducial} which we set equal to the fiducial parameters given there. The assumptions we make for the various observables we consider are defined in what follows. 

\begin{table}[b!]
\begin{center}
 \begin{tabular}{|l |l |} 
 \toprule
   Parameter & Fiducial Value             \\  
 \hline
Cold dark matter density ($\Omega_c h^2$) &  0.1197  \\ Baryon density ($\Omega_b h^2$)  &   0.0222 \\
Angle subtended by acoustic scale ($\theta_s$)  & 0.010409 \\
Optical depth to recombination $(\tau)$ &   0.060 \\
Primordial scalar fluctuation amplitude ($A_s$)  &   $2.196\!\times\!10^{-9}$  \\
Primordial scalar fluctuation slope ($n_s$) &  0.9655	    \\
  \hline
\end{tabular}
    \caption{
    Fiducial cosmological parameters for the 6-parameter $\Lambda$CDM model we consider in our calculations throughout this paper.}
\label{table:cosmo_fiducial}
\end{center}
\end{table}

\vspace*{-0.25cm}
\subsection{CMB}
\vspace*{-0.25cm}

We model the instrumental and atmospheric noise contributions to the CMB temperature as 
\be\label{eq:instrument_noise1}
N_\ell^{\rm TT}=\Delta_T^2\exp\left(\frac{\ell(\ell+1)\theta^2_{\rm FWHM}}{8\ln2}\right)\left[1+\left(\frac{\ell}{\ell_{\rm knee}}\right)^{\alpha_{\rm knee}}\right]\,,\non
\ee
where $\Delta_T$ is the detector RMS noise and $\theta_{\rm FWHM}$ is the Gaussian beam full width at half maximum. The second term inside the brackets in Eq.~\eqref{eq:instrument_noise1} corresponds to the `red' noise due to Earth's atmosphere, parametrised by the terms $\ell_{\rm knee}$ and $\alpha_{\rm knee}$. We define our choices for these parameters to match the ongoing and upcoming CMB surveys in Table~\ref{tab:beamnoise}.

\begin{table}[t!]
\begin{tabular}{|l|c|c|c|c|}
    \hline & \multicolumn{2}{c|}{Beam FWHM} & \multicolumn{2}{c|}{Noise RMS~$\mu$K'} \\ 
    \cline{2-5} 
    & \multicolumn{1}{c|}{SO} & \multicolumn{1}{c|}{CMB-S4} &  \multicolumn{1}{c|}{SO} & 
    \multicolumn{1}{c|}{CMB-S4} \\ \hline
    39 GHz  & $5.1'$ & $5.1'$ & 36 & 12.4  \\
    93 GHz  & $2.2'$ & $2.2'$ & 8 & 2.0 \\
    145 GHz & $1.4'$ & $1.4'$ & 10 & 2.0 \\
    225 GHz & $1.0'$ & $1.0'$ & 22 & 6.9 \\
    280 GHz & $0.9'$ & $0.9'$ & 54 & 16.7 \\ \hline
\end{tabular}
\vspace{-0.3cm}
\caption{{\it Inputs to ILC noise: } The beam and noise RMS parameters we assume for survey configurations roughly corresponding to Simons Observatory (SO) and CMB-S4. In all cases, we account for the degradation due to Earth's atmosphere by defining the CMB noise choose $\ell_{\rm knee}=100$ and $\alpha_{\rm knee}=-3$.}
\vspace*{-0.4cm}
\label{tab:beamnoise}
\end{table}

The millimeter-wavelength CMB signal gets contributions also from the black-body late-time and reionization kSZ, the Poisson and clustered CIB, as well as the tSZ foregrounds, which we calculate following Refs.~\citep{Madhavacheril:2017onh,Park:2013mv}. We omit the cross-correlation between tSZ and CIB. We include radio sources following Ref.~\citep{Lagache19}. We calculate the lensed CMB black-body using CAMB~\citep{CAMB}.

\vspace*{-0.25cm}
\subsection{Galaxy surveys}
\vspace*{-0.25cm}

{Galaxy surveys play a significant role in detecting velocity fluctuations upon cross correlation with reconstructed radial-velocity fields from kSZ tomography. They also serve as small-scale tracers of the electron density and can be used for velocity reconstruction through cross correlation with CMB on small scales. As these programs will be pursued in the near future, we include in our analysis the galaxy density fields anticipated to be observed in the near future.}

{We consider the ongoing measurements of quasi-stellar objects (QSOs) with DESI~\citep{2016arXiv161100036D} and {high-redshift galaxies with photometric LSST survey~\citep{2009arXiv0912.0201L}}. We follow Ref.~\citep{2016arXiv161100036D} for DESI quasars when calculating the number density and set the bias to satisfy $b_g(z)=1.2/D(z)$. For LSST, we approximate the galaxy density of the ``gold'' sample, with $n_\text{gal}(z) = n_0[({z}/{z_0}]^2\exp(-z/z_0)/{2z_0}$ with $n_0=40~\text{arcmin}^{-2}$ and $z_0=0.3$ and take the galaxy bias as $b_g(z)=0.95/D(z)$. For LSST, we consider the standard anticipated photo-$z$ error $\sigma_z=0.03(1+z)$ which becomes increasingly more detrimental at higher redshifts. For DESI, the photo-$z$ errors will be small $\sigma_z\ll1$. We show the galaxy bias and number density that we consider for these surveys for a range of redshifts in Table~\ref{tab:survey-specs} for reference.}
\begin{table}[t!]
  \begin{center}
    \begin{tabular}{| l | l l l l | c |} 
    \hline
     LSST & $\,\!z\!=1.9$ & 2.6 & 3.45 & 4.45 &  \\  
     $b_g$ & 1.81 &  2.47 & 3.28 & 4.23 & \\
     $n_{\rm gal}$ ($\times10^{4}$) $[{\rm Mpc}^{-3}]$ & 14.9 &  2.9 & 0.34 & 0.02  & Y10\\
     \hline
     DESI & & & &  & \\ 
     $b_g$ & 1.92 & 3.18 & 4.71 & 6.51  & \\
     $n_{\rm gal}$ ($\times10^{6}$) $[{\rm Mpc}^{-3}]$& 1.61 &  0.80 & 0.15 & 0.03 &  $r<22.5$\\
     $n_{\rm gal}$ ($\times10^{6}$) $[{\rm Mpc}^{-3}]$& 2.20 &  1.18 & 0.30 & 0.04 &  $r<23.0$\\
    \hline
   \hline
   \hline
   \end{tabular}
   \vspace{-0.2cm}
   \caption{{\textit{Assumed galaxy bias $b_g$ and number density $n_{\rm gal}$ for DESI and LSST at various redshifts.} We consider two choices for the expected distribution of QSO redshifts from DESI following Ref.~\citep{2016arXiv161100036D} with two different quasar luminosity function $r$ smaller than 22.5 and 23. For the LSST survey we take the anticipated number counts after 10 years of observations.}}
    \label{tab:survey-specs}
  \end{center}
  \vspace{-1cm}
\end{table}

\begin{figure*}[t!]
    \centering
    \includegraphics[width =\linewidth]{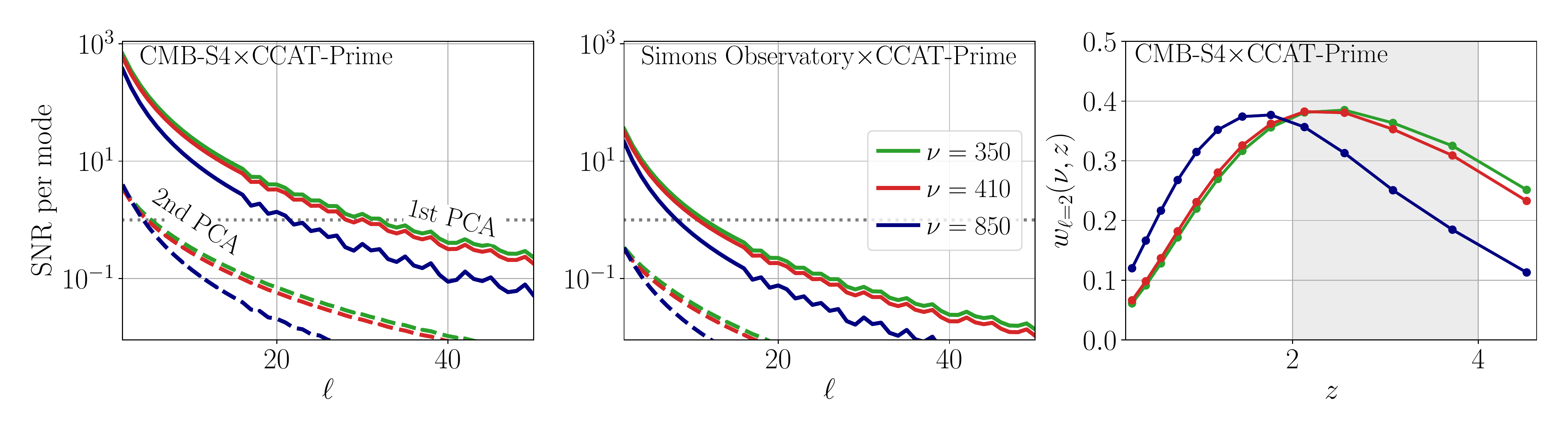}
    \vspace*{-1cm}
    \caption{The left and center panels show the anticipated SNR per mode from the first (second) principle components corresponding to the CCAT-Prime measurements, shown with solid (dashed) lines. In both panels we show results for measurements  at frequencies $\nu=\{350,410,850\}$ GHz. The left (center) panel corresponds to using anticipated CMB noise levels matching CMB-S4  (Simons Observatory) for kSZ tomography (velocity reconstruction). The right panel shows the redshift weights of the first principle component for $\ell=2$, normalised to satisfy $\sum_\alpha w_2(\nu,z_\alpha)=1$ for $\alpha=\{1,\ldots,N_{\rm bin}\}$. Here, we only show the results for correlations of CIB with CMB-S4-like CMB maps; since results from using a Simons Observatory-like CMB experiment are largely identical. For all panels we set the number of redshift bins $N_{\rm bin}$ equal to 13. Increasing $N_{\rm bin}$ does not improve the results as both the signal and the reconstruction noise are highly correlated between different redshift bins.}
    \label{fig:SNR_PCA}
    \vspace*{-0.5cm}
\end{figure*}

\vspace*{-0.25cm}
\subsection{The CIB noise and detection}
\vspace*{-0.25cm}

The current state-of-the-art measurements of CIB are provided by the \textit{Planck} satellite at frequencies 216, 353, 545 and 857 GHz. These provide high-fidelity CIB maps within the multipole range of $200\gtrsim L\gtrsim 2500$. Similar to Ref.~\citep{McCarthy:2019xwk}, however, we also find that the prospects to reconstruct the large-scale radial velocity field from \textit{Planck} CIB maps is not optimistic due to \textit{Planck}'s $\sim\!5$ arc-minute resolution and the residual extra-galactic foregrounds which are significant even after ILC-cleaning. The picture is much more optimistic, however, for the upcoming CCAT-Prime survey, which will make high-resolution measurements of the CIB at a range of frequency bands including 350, 410 and 850 GHz~\citep{CCAT-Prime:2021lly}. 

We define the CIB instrumental noise term as 
\be\label{eq:instrument_noise}
N_\ell^{I_\nu I_\nu}=\Delta_T^2\exp\left(\frac{\ell(\ell+1)\theta^2_{\rm FWHM}}{8\ln2}\right)+\Delta_R^2\left(\frac{\ell}{\ell_{\rm knee}}\right)^{\alpha_{\rm knee}}\!\!\!\!\!\!\!\!\!\!,\,\,\,\,\,\,
\ee
and set $\Delta^2_R=1000$ and $\alpha_{\rm knee}=-3.5$ to match anticipated measurements of CCAT-Prime~\citep{CCAT-Prime:2021lly}. Our choices for the instrumental noise of the \textit{Planck} and CCAT-prime surveys are shown in Table~\ref{tab:noise_1}. The anticipated signal and noise spectra matching CCAT-Prime specifications were shown in Fig.~\ref{fig:signal_CIB_lensing}.

\begin{table}[t]
\begin{tabular}{|l|c|c|c|c|c|c|}
\hline
& \multicolumn{3}{c|}{\textit{Planck}} & \multicolumn{3}{c|}{CCAT-Prime} \\
\hline
Frequencies (GHz)
& \multicolumn{1}{c|}{353} & \multicolumn{1}{c|}{545} &
\multicolumn{1}{c|}{857} & \multicolumn{1}{c|}{350} &
\multicolumn{1}{c|}{410} &
\multicolumn{1}{c|}{850}\\ \hline
$\theta_{\rm FWHM}$ & $4.94'$ & $4.83'$ & $4.64'$ & $37''$ & $32''$ & $15''$ \\
$\Delta_T (\rm \mu K\!\!-\!\!\rm arcmin)$ & $0.036$ & $0.20$ & $4.85$ & 107 & 407 & $6.8\!\times\!10^{5}$ \\
\hline
\end{tabular}
\caption{Inputs to instrumental noise parameters defined in Eq.~\eqref{eq:instrument_noise} matching the specifications of \textit{Planck} and the upcoming CCAT-Prime. We account for the `red' noise due to Earth's atmosphere on the latter measurement by setting $\ell_{\rm knee}=100$, $\alpha_{\rm knee}=-3.5$ and $\Delta_R^2=1000$.}
\label{tab:noise_1}
\vspace*{-0.5cm}
\end{table}

In order to build intuition on the information contained in the reconstructed velocity field, we first perform a principal component anaylsis (PCA). We calculate a diagonal matrix with entries equal to the signal-to-noise of each principle component via the Karhunen-Loeve technique as 
\be
\boldsymbol{C}^{\nu\rm PCA}_\ell= \boldsymbol{R}^\nu_3 \boldsymbol{R}^\nu_2\boldsymbol{R}^\nu_1 \boldsymbol{S}_\ell \boldsymbol{R}^{\nu,T}_1 \boldsymbol{R}^{\nu,T}_2 \boldsymbol{R}^{\nu,T}_3\,,
\ee
which consists of three rotations\footnote{Note we have omitted the multipole $\ell$ dependence of the rotation matrices here for brevity.}: \textit{(1.)}~$\boldsymbol{R}^\nu_1$ diagonalizes the noise covariance matrix $\boldsymbol{N}^\nu_\ell$ at a given measurement frequency $\nu$ and multipole $\ell$. \textit{(2.)} $\boldsymbol{R}^\nu_2$ sets the transformed noise matrix ${\boldsymbol{N}^\nu_\ell}' = \boldsymbol{R}^\nu_1 \boldsymbol{N}^\nu_\ell \boldsymbol{R}_1^{\nu,T}$ to identity, i.e. ${\boldsymbol{N}^\nu_\ell}'=\boldsymbol{I}$, and \textit{(3.)} $\boldsymbol{R}_3$ diagonalizes the matrix $\boldsymbol{R}_2 \boldsymbol{R}_1 \boldsymbol{S}_\ell \boldsymbol{R}_1^{\nu,T} \boldsymbol{R}_2^{\nu,T}$. Here, all matrices are $N_{\rm bin}\times N_{\rm bin}$ square matrices where $N_{\rm bin}$ is the number of redshift bins considered in our analysis, $T$ superscripts indicates matrix transpose and $\boldsymbol{S}_\ell$ is the covariance of the radial velocity field signal satisfying $(\boldsymbol{S}_\ell)_{\alpha\beta}=C_{\ell,\alpha\beta}^{\bar{v}\bar{v}}$. In order to calculate the detection significance, we omit the cosmic variance ($\boldsymbol{S}_\ell$) from the noise covariance, setting $(\boldsymbol{N}^\nu_\ell)_{\alpha\beta} = N_{\ell,\alpha\beta\nu}^{\bar{v}\bar{v}}$ defined in Eq.~\eqref{eq:CIB_kSZ_unbiased}. 

The resulting $\boldsymbol{C}^{\nu\rm PCA}_\ell$ is a diagonal matrix whose entries correspond to the (detection) signal-to-noise ratio (SNR) of the $N_{\rm bin}$ principle components for a given velocity mode. We show the detection SNR from the first and second principle components of the velocities anticipated to be reconstructed from the CCAT-Prime survey measurements on the left panel of Fig.~\ref{fig:SNR_PCA}. The shape of the first principle component in the redshift basis for multipole $\ell=2$ is shown on the right panel of the same figure. It is valuable to note that the contribution to the lower-frequency CIB signal comes largely from redshifts $2<z<4$; the anticipated period of helium reionization. 

\vspace*{-0.25cm}
\subsection{Weak lensing}
\vspace*{-0.25cm}

In order to fully capitalise on the high-redshift probes of large-scale structure, we also consider lensing reconstruction from measurements of CMB temperature and polarization. We calculate the minimum-variance noise on the reconstructed lensing deflection using the standard quadratic estimator from Ref.~\citep{Okamoto:2003zw}, which we described in Appendix~B of Ref.~\citep{Hotinli:2021umk}. We take the CMB polarization noise spectra to satisfy $N_{\ell}^{EE} = N_{\ell}^{BB} = 2N_{\ell}^{TT}$, as is expected with fully-polarized detectors, and set the  maximum multipole used in lensing reconstruction $\ell_{\rm max}$ equal to $5000~(7000)$ for reconstructions including the temperature signal~(only polarization signals). Our choices for the parameters describing CMB measurement noise are given in Table~\ref{tab:beamnoise}. The lensing reconstruction noise and the lensing power spectrum are shown on the right panel of Fig.~\ref{fig:signal_CIB_lensing}. We use \texttt{class\_delens}\footnote{Publicly available at \href{https://github.com/selimhotinli/class_delens}{github/selimhotinli/class\_delens}.} code for our calculations which provides a self-consistent, iterative, all-orders treatment of CMB delensing and lensing-noise reconstruction on the curved sky, as described in Ref.~\citep{Hotinli:2021umk}.

\section{Probing helium reionization}\label{sec:helium}

As an application of the high-redshift large-scale-structure probes and velocity reconstruction, here we evaluate the prospects to probe helium reionization with velocity tomography. Similar to Ref.~\citep{Hotinli:2022jna}, we characterise the change in the ionization fraction during helium reionization with a hyperbolic tangent
\be\label{eq:mean_reio}
\overline{x}_e(z)\!=\!\frac{1}{2}\!\left[2\!+\!\Delta \bar{x}_{\rm He}\!-\!\Delta \bar{x}_{\rm He}\tanh{\left(\frac{y(z_{\rm re}^{\rm He})\!-\!y(z)}{\Delta_y^{\rm He}}\right)}\right]\!,\,\,\,\,\,
\ee
where $\Delta \bar{x}_{\rm He}$ determines the total change in the mean ionization fraction during helium reionization, $z_{\rm re}^{\rm He}$ is the redshift half-way through the helium reionization, $y(z)=(1+z)^{3/2}$, and $\Delta_y^{\rm He}$ parameterizes the duration of the transition. In what follows we will replace $\Delta\bar{x}_{\rm He}$ with $Y_p$ and the $\Delta_y^{\rm He}$ parameter with $\Delta_z^{\rm He}$, which we define as the duration in redshift of the central 50$\%$ change in ionization fraction. {We use CAMB to calculate $\partial Y_p/\partial \Delta \bar{x}_{\rm He}$.} 

The reconstructed velocities depend on the free electron fraction through the optical depth. If helium reionization has not been modelled correctly to match the data, the resulting velocity reconstruction will be biased. We find to a good approximation the reconstructed velocities from kSZ tomography using both galaxies and CIB satisfy $\hat{\bar{v}}^\alpha(\nhat)\simeq[{\bar{x}_e(z_\alpha)}/{\bar{x}_e(z_\alpha)_{\rm fid}}]b_X(z_\alpha){\bar{v}}^\alpha(\nhat)\,,$ where $\bar{x}_e(z)/{\bar{x}_e(z)}_{\rm fid}$ is equal to unity if the true helium reionization match the fiducial model, and $b_X(z_\alpha)$ is the standard kSZ optical-depth bias due to mismodelling of the cross power of electron and some tracer of large-scale structure $X$ at small scales, as described in e.g.~Refs.~\citep{Munchmeyer:2018eey, Smith:2018bpn,Deutsch:2017ybc}.

\begin{figure}[t!]
    \centering
    \includegraphics[width = 1.0\linewidth]{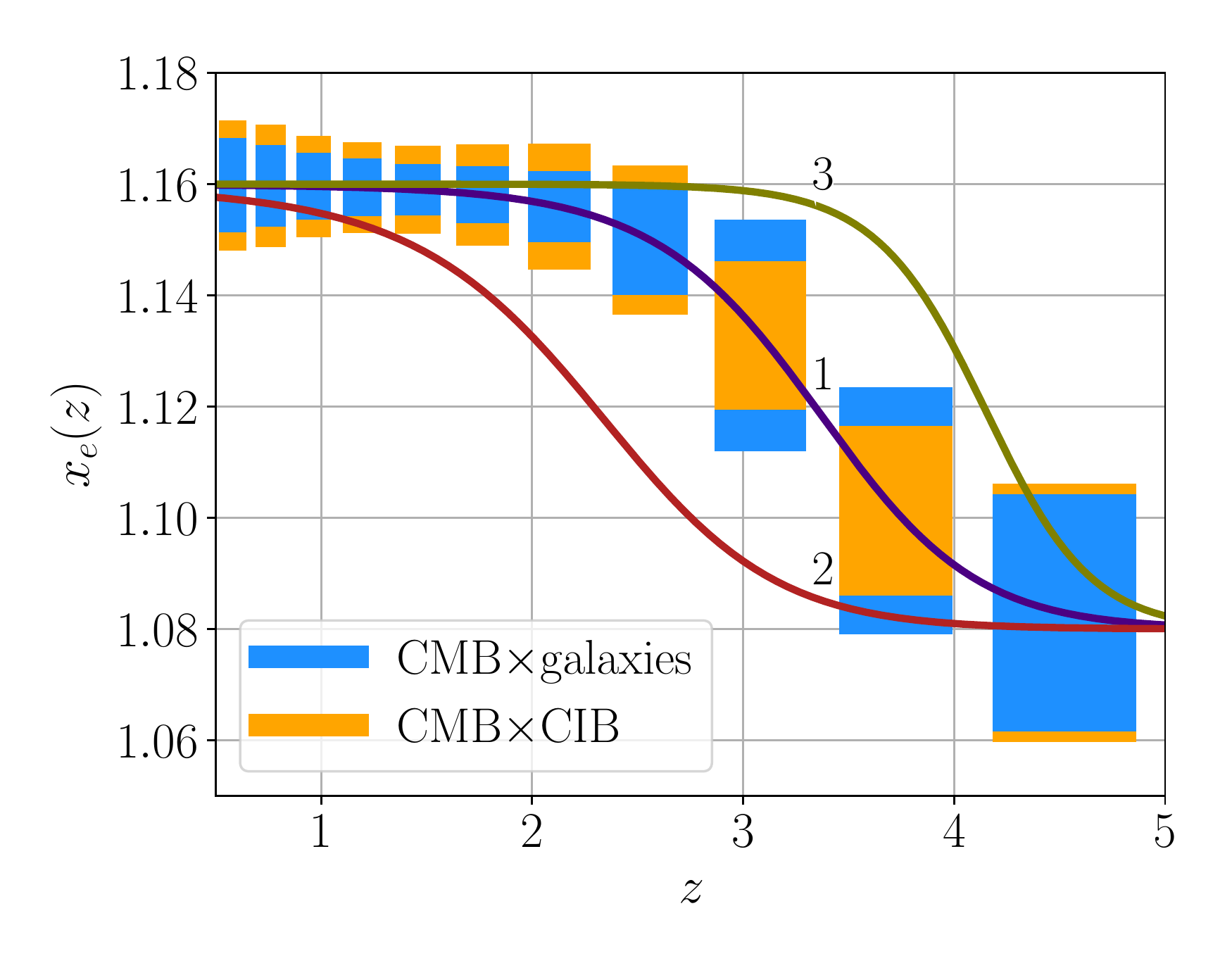}
    \vspace*{-0.75cm}
    \caption{The $1\sigma$ errors on the amplitudes of reconstructed velocity fields from combination of CMB and galaxies (CIB) shown with blue (orange) error bars. In both cases we also include the cross-correlation of the reconstructed large-scale  velocity with the galaxy density, latter anticipated to match LSST observations. The coloured solid lines labelled 1 to 3 correspond to three distinct helium reionization models with varying fiducial values and considerations described in the text.}
    \vspace*{-0.5cm}
    \label{fig:bias_constraints}
\end{figure}

In order to measure the information
content of the correlated reconstructed velocity and density observables, we define an ensemble information matrix as
\be\label{eq:fisher}
&&\mathcal{F}_{ik}\\ 
&&\!\!\!=\!\!\!\!\sum\limits_{\ell=\ell_{\rm min}}^{\ell_{\rm max}}\!\!\!f_{\rm sky}\frac{2\ell+1}{2}{\rm Tr}\! \left[\frac{\partial \boldsymbol{S}_\ell}{\partial \pi_i}(\boldsymbol{S}_\ell\! +\!\boldsymbol{N}_\ell)^{-1}\frac{\partial \boldsymbol{S}_\ell}{\partial \pi_k}(\boldsymbol{S}_\ell\!
+\!\boldsymbol{N}_\ell)^{-1}\right]\nonumber
\ee
where $\boldsymbol{S}_\ell$ $(\boldsymbol{N}_\ell)$ is the signal (noise) matrix and $\partial\boldsymbol{S}_\ell/\partial\pi_i$ represents the derivative of the signal matrix with respect to parameter $\pi_i$. Throughout this paper we set $f_{\rm sky}$ equal to 0.4 (0.3) to match forecasts including the anticipated CMB-S4 (Simons Observatory) measurements. Similarly, we assume the joint sky coverages of large-scale structure tracers we consider here (measurements of CCAT-Prime and galaxy surveys) with CMB-S4 and Simons Observatory are 0.4 and 0.3, respectively. We set $\ell_{\rm max}=200$ and $\ell_{\rm min}=2$ throughout, unless specified otherwise. In addition to the parameters characterising helium reionization, we consider bias parameters for the galaxy, velocity and lensing observables, as well as the amplitude of primordial scalar perturbations $A_s$ as free parameters in our forecasts.

We demonstrate the measurement accuracy of the velocity reconstruction from kSZ tomography using galaxies (blue error bars) and CIB (orange error bars) in Fig.~\ref{fig:bias_constraints}. The error bars in this figure correspond to $1\sigma$ errors on the amplitudes of the reconstructed velocity fields in 13 redshift bins, which we define as $\hat{\bar{v}}^{\alpha}_{\ell m}=b_\alpha \bar{v}^{\alpha}_{\ell m}$. Here, the information matrix consists of \textit{(1.)} the covariance of velocity fields reconstructed at each redshift bin, $C_{\ell,\alpha\beta}^{\hat{\bar{v}}\hat{\bar{v}}}=C_{\ell,\alpha\beta}^{{\bar{v}}{\bar{v}}} +N_{\alpha\beta\ell}^{{\bar{v}}{\bar{v}}}$, where $N_{\alpha\beta\ell}^{{\bar{v}}{\bar{v}}}$ is the reconstruction noise defined in Eq.~\eqref{eq:noise_rec_CIB} for CMB$\times$CIB tomography and in Eq.~\eqref{eq:noise_kSZ_gal} for CMB$\times$galaxy tomography and $C_{\ell,\alpha\beta}^{{\bar{v}}{\bar{v}}}$ is the redshift-binned radial velocity power spectra; \textit{(2.)} the cross-correlation between the reconstructed velocity and galaxy fields, $C_{\ell,\alpha\beta}^{{{\bar{v}}}\delta_g}$; as well as \textit{(3.)} the covariance of the observed large-scale galaxy fields $C_{\ell,\alpha\beta}^{\delta_g\delta_g,\rm obs}$, which includes the galaxy shot noise. 

For kSZ tomography using CIB, we consider three frequencies at $\{350,410,850\}$\,GHz and experimental specifications matching CCAT-Prime. The information matrix in this case has the shape $(4N_{\rm bin}\times4N_{\rm bin})$ and includes the cross-correlation between reconstructed velocities at different frequencies and redshifts. For kSZ tomography using galaxies, the information matrix has the shape $(2N_{\rm bin}\times2N_{\rm bin})$. In both cases we forecast assuming CMB-S4 and LSST. In addition to the velocity bias parameters we defined above, here we also marginalise over the galaxy biases and the three reionization parameters.

The solid lines labeled with numbers 1 to 3 correspond to three helium reionization models with fiducial choices for ($z_{\rm re}^{\rm He},\Delta_z^{\rm He}$) set equal to  $(3.34,0.8)$, $(2.29,0.79)$, and $(4.14,0.58)$, respectively. We take $Y_p=0.245~(\Delta x_{\rm He}\simeq0.08)$ for all models. As done in Ref.~\citep{Hotinli:2022jna}, these models are chosen to roughly match models H1, H3 and H6, considered in Ref.~\citep{LaPlante:2016bzu}, respectively, which represent several plausible and distinct models of helium reionization.\footnote{Model H1 reproduces the quasar spectrum measured by Ref.~\citep{Lusso2015}, the quasar abundance measured by Refs.~\citep{Ross2013, Masters2012, McGreer2013}, and quasar clustering measured by BOSS \citep{White2012}. Model H3 considers a quasar abundance that is reduced by a factor of 2. This model is consistent with the measured uncertainties but yields a slightly delayed reionization scenario. Model H6 reproduces the semi-numeric models of Ref.~\citep{Haardt2012} and uses a uniform UV background rather than explicit quasar sources.} Distinguishing between these models can provide an independent determination of the average luminosity and abundance of quasars and their interactions with the IGM, which complements direct measurements from spectroscopic surveys.

\begin{figure}[t!]
    \centering
    \includegraphics[width = 1.0\linewidth]{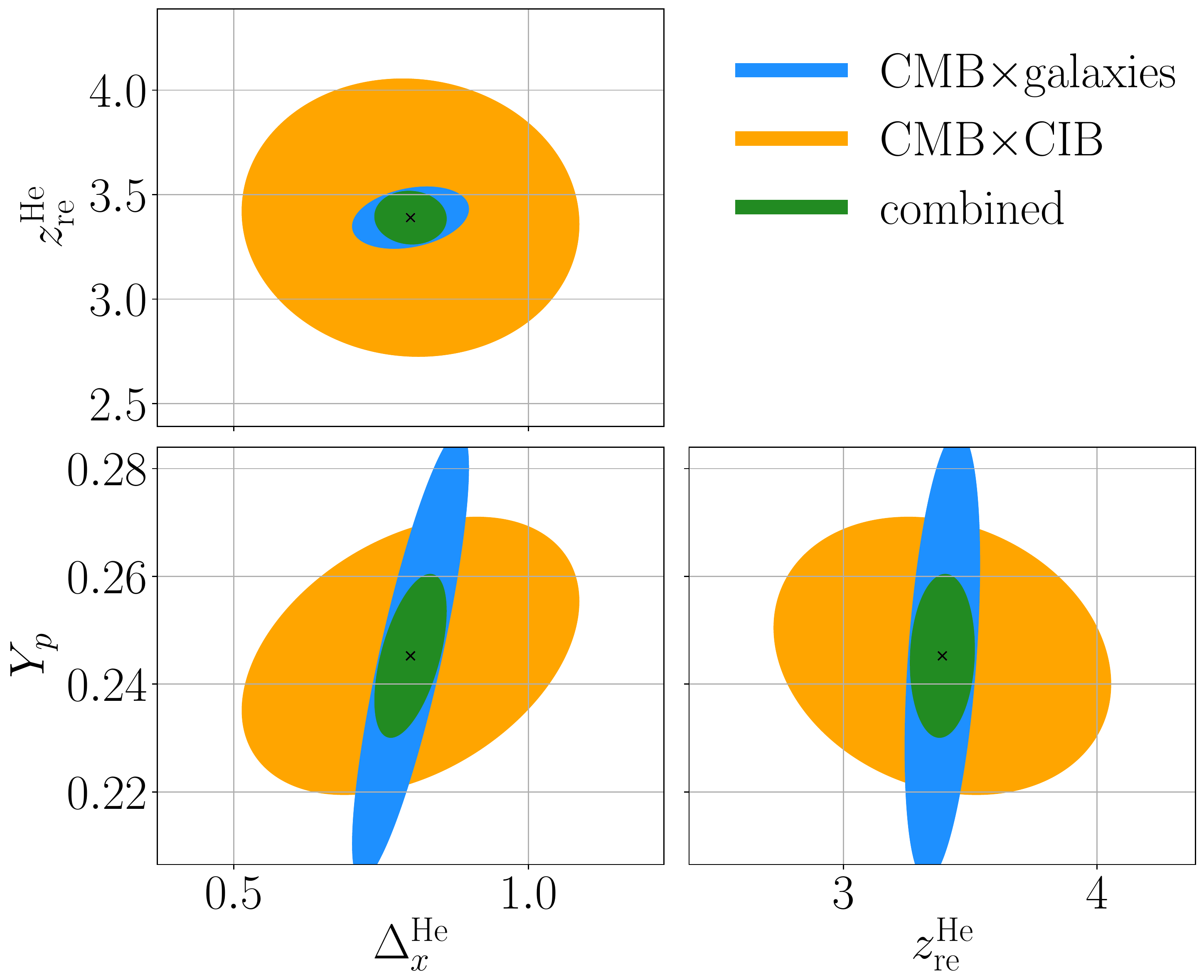}
    \vspace*{-0.75cm}
    \caption{The $1\sigma$ error contours on helium reionization model parameters defined in the text. Similar to Fig.~\ref{fig:bias_constraints}, blue (orange) contours correspond to reconstructed velocity fields from combination of CMB and galaxies (CIB). }
    \vspace*{-0.45cm}
    \label{fig:fisher_constraints}
\end{figure}

Next, we forecasts the measurement accuracy of helium reionization model parameters assuming model 1 in Fig.~\ref{fig:fisher_constraints}. Similar Fig.~\ref{fig:bias_constraints}, we consider a joint analysis of small-scale CMB measurements matching the anticipated noise and foregrounds of the upcoming CMB-S4 survey; galaxy surveys with specifications matching the upcoming LSST survey; and CIB measurements that are anticipated to match the upcoming CCAT-Prime survey. The blue (orange) contours correspond to $1\sigma$ measurement errors on helium reionization parameters from measurements of the galaxy and velocity fields on large scales, latter reconstructed from combination of CMB and galaxies (CIB) on small scales. The green contours correspond to the combined measurement accuracy anticipated from these observables. 

For the results in Fig~\ref{fig:fisher_constraints}, we have defined a parameterised velocity bias in the form $b^X(z)=b_0^X+b^X_1 z+ b^X_2 z^2$, where $X=\{\rm CIB,\rm gal\}$, separately for the velocity reconstructions using CIB and galaxies. We set the fiducial values of $\{b_0^{\rm CIB}, b_1^{\rm CIB}, b_2^{\rm CIB}\}$ to $\{0.84, 0.74, 0.32\}$ following Refs~\citep{Maniyar:2018xfk,McCarthy:2020qjf}. For the kSZ optical depth bias from kSZ using galaxies, we set the fiducial value $b_0^{\rm gal}$ to unity and $\{b_1^{\rm gal},b_2^{\rm gal}\}$ to zero. We also define the bias on the large-scale galaxy density with the same parametrization. When inferring errors on helium reionization model parameters, we marginalize over these biases as well as the amplitude of scalar primordial fluctuations $A_s$. Unless otherwise specified, we also assume $10\%$ priors on the bias parameters $b_0^{\rm CIB}$ and $b_0^{\rm gal}$, latter can be potentially achieved by measurements of the CMB polarization~\citep[e.g.][]{Lee:2022udm,Hotinli:2022wbk}, the moving lens effect~\citep{Hotinli:2021hih,Hotinli:2020ntd,Hotinli:2018yyc} or the fast radio bursts~\citep{Madhavacheril:2019buy}. Our results for kSZ tomography using LSST galaxies match  our earlier work~\citep{Hotinli:2022jna} within a factor of $\sim2$, although note Ref.~\citep{Hotinli:2022jna} used a 3-dimensional box formalism. For the experimental specifications matching CCAT-Prime and LSST, we find kSZ tomography with these tracers provide comparable information on the helium reionization.

\begin{table}[t!]
  \begin{center}
    \label{tab:detectionSNR}
    \begin{tabular}{| c | c | c |} 
    \hline
    Detection SNR &   \multicolumn{2}{c|}{Experiments} \\ 
    \hline
    Observables & SO \& DESI & CMB-S4 \& LSST (GS) \\
   \hline
   $\hat{\bar{v}}^\alpha_{\rm CIB},\, \delta_g^\alpha$ & 0.8 & 3.9  \\
   $\hat{\bar{v}}^\alpha_{\rm gal},\, \delta_g^\alpha$ & 0.5 &
   2.8 \\
   $\hat{\bar{v}}^\alpha_{\rm CIB},\hat{\bar{v}}^\alpha_{\rm gal},\, \delta_g^\alpha$ & 1.2 & 6.5 \\ 
   $\hat{\bar{v}}^\alpha_{\rm CIB},\hat{\bar{v}}^\alpha_{\rm gal},\, \delta_g^\alpha, \hat{\phi}$ & 1.6 & 6.9 \\
   \hline
   \end{tabular}
   \caption{Detection SNR of helium reionization defined as the $1\sigma$ measurement error on the $\Delta \bar{x}_{\rm He}$ parameter after marginalising over other reionization parameters and biases as defined in the text. Here, $\hat{\bar{v}}^\alpha_{\rm CIB}$ refers to the radial velocity field reconstructed from small-scale CMB$\times$CIB cross-correlation. We use CIB measurements with anticipated experimental specifications matching CCAT-Prime throughout. The velocity fields reconstructed from cross-correlation of CMB and galaxies are shown with $\hat{\bar{v}}^\alpha_{\rm gal}$. Here, $\delta_g^\alpha$ refers to the large-scale galaxy field anticipated to be observed with either DESI or LSST gold sample (GS). The lensing potential reconstructed from CMB temperature and polarization fluctuations is shown with $\hat{\phi}$.}
  \end{center}
  \vspace{-0.8cm}
\end{table}

In order to assess the detection prospects of helium reionization, we consider the signal-to-noise (SNR) on $\Delta \bar{x}_e$ after marginalising over other reionization and bias parameters. For velocity reconstruction using CIB, we consider the CCAT-Prime specifications throughout. We find that the combination of reconstructed velocities from Simons Observatory and CIB or DESI galaxies will likely not reach sufficient SNR to detect helium reionization when considered in isolation. Nevertheless, we find that hints of helium reionization may be detected at $\sim1-2\sigma$ from the joint analysis of these signals and weak lensing. For CMB-S4 and an LSST-like survey assuming specifications matching the `gold sample', we find helium reionization may be detected at around $\sim2-4\sigma$ from CIB- and galaxy-reconstructed velocity fields in isolation; and that the detection SNR can reach $\sim6-8\sigma$ if these signals are jointly analysed. We show the detection SNR of helium reionization from different considerations in Table~\ref{tab:detectionSNR}. Including anticipated high-redshift galaxy dropouts following Refs.~\citep{Ferraro:2022twg, 2018PASJ...70S..10O, Harikane:2017lcw} increase the prospects of detecting helium reionization with LSST by over $\sim30-40$ percent using the methods we consider here.  

\begin{figure}[t!]
    \centering
    \includegraphics[width = 1.0\linewidth]{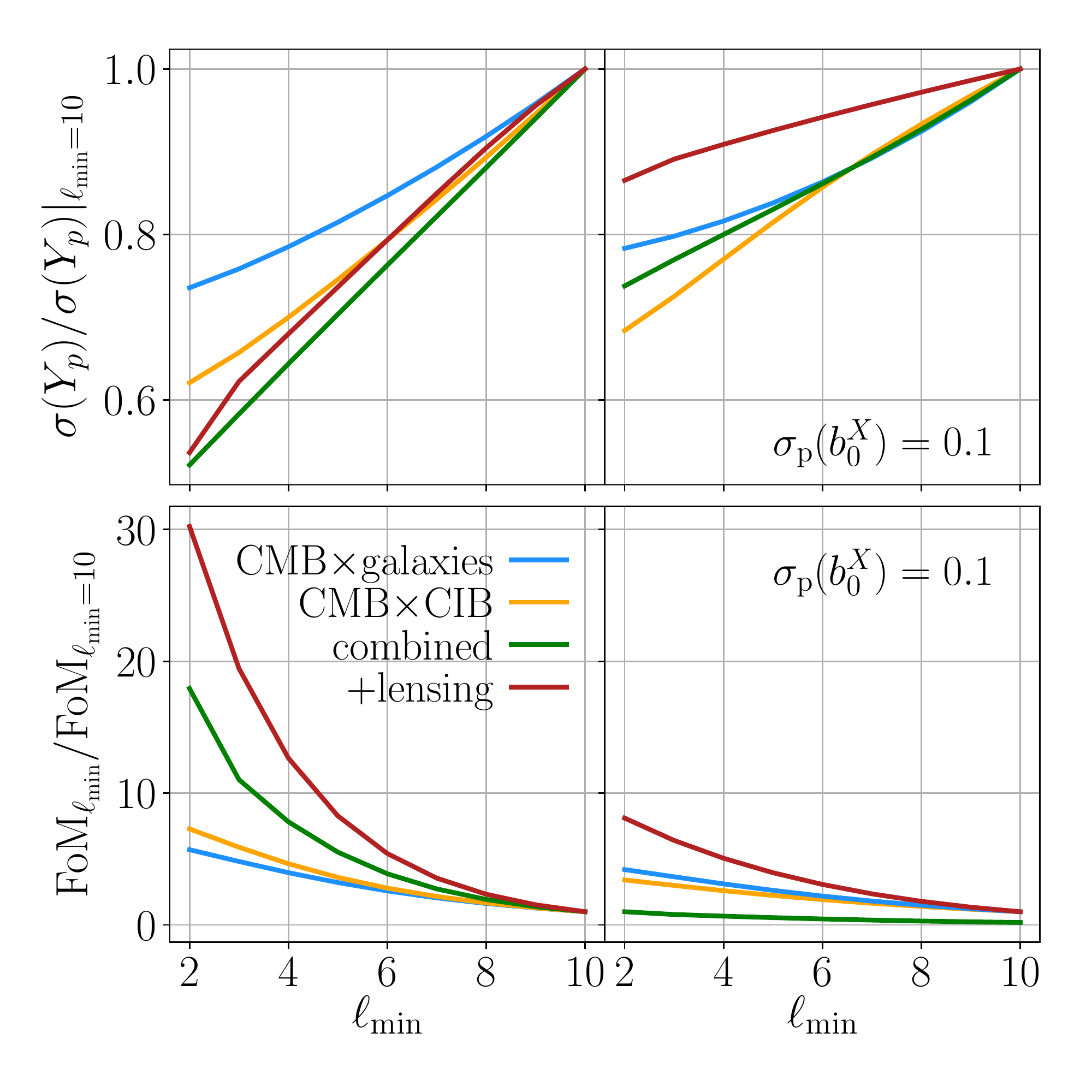}
    \vspace*{-1.1cm}
    \caption{Sensitivity of the statistical power of kSZ tomography to largest scales considered in the analysis. Upper panels show the sensitivity of measurement of the primordial helium abundance $Y_p$, parameterised as the ratio between the error on $Y_p$ for a choice of $\ell_{\rm min}$ and when $\ell_{\rm min}$ is set to 10. The lower panels correspond to fractional improvement in the Figure of Merit (FoM) as a function of $\ell_{\rm min}$. The right panels correspond to assuming 10 percent priors on the velocity  bias parameters. The solid blue lines correspond to kSZ tomography using LSST galaxies as described in the text. The orange lines correspond to kSZ tomography using CIB. Green lines correspond to combination of these obserables. The red lines also include the reconstructed lensing potential from measurement of CMB weak lensing.}
    \label{fig:lmin_analyse}
    \vspace*{-0.5cm}
\end{figure}

The prospect of measuring cosmological signatures at high redshifts using kSZ tomography depends significantly on the lowest accessible multipoles (largest angular scales) at which the velocity fields can be reconstructed. In Fig.~\ref{fig:lmin_analyse} we demonstrate the dependence of the helium reionization detection as well as the Figure of Merit (FoM) on the minimum multipole we consider in our forecasts $\ell_{\rm min}$. The FoM provides a simple quantitative summary of how well a given observable can improve the prospects of measuring cosmological signatures at high-redshifts, and is defined as ${\rm FoM}=\sqrt{\det{\mathcal{F}_{ij}^{-1}}}$ where $\mathcal{F}_{ij}^{-1}$ is the information matrix defined in Eq.~\eqref{eq:fisher}. We find the FoM improves by a factor of $\sim5$ for the reconstructed velocities using CIB and galaxies in isolation, and by a factor $\sim15$ when these observables are jointly analysed. We find the improvement of the FoM is more enhanced when lensing reconstruction is also considered in the analysis. These results are shown on the lower panels of Fig.~\ref{fig:lmin_analyse}. The lower-right panel corresponds to assuming no priors on the reconstructed velocity, while  the lower-left panel assumes $10$ percent priors satisfying $\sigma_p(b_0^{X})=0.1$ where $X=\{\rm CIB,gal\}$. The reduced sensitivity of the statistical power of observables when we consider priors on these biases suggests a significant  portion of the statistical power on large scales contribute to constraining these parameters. 

The upper panels on Fig.~\ref{fig:lmin_analyse} correspond to fractional improvement on the errors on $Y_p$ compared to assuming $\ell_{\rm min}=10$. The increase in error from increasing $\ell_{\rm min}$ can be seen to reach a factor $\sim2$ by $\ell_{\rm min}\sim6$ if no priors are assumed on the velocity reconstruction. Similar to the FoM, assuming 10 percent priors on these parameters lowers the sensitivity of the measurement accuracy to $\ell_{\rm min}$. Overall, we note that our ability to reconstruct largest angles will play a crucial role in realising the prospects of cosmological inference at high redshifts including detecting and characterising helium reionization using kSZ tomography. 

\vspace*{-0.5cm}
\section{Discussion}\label{sec:discussion}
\vspace*{-0.25cm}

Our results suggest detecting and characterising helium reionization in the next 1-3 years via kSZ tomography using upcoming surveys such as Simons Observatory and DESI may be difficult. However we note that the forecasts we included involving DESI are likely somewhat pessimistic given the choice of the low number of redshift bins we considered for this experiment. As the photometric redshift errors will be very small for DESI, we would expect increasing the number of redshift bins would increase the information content of the reconstructed velocities and galaxy-velocity cross-correlation for kSZ tomography using DESI galaxies in principle. Moreover, futuristic experiments such as the proposed MegaMapper~\citep{Schlegel:2019eqc, Ferraro:2022cmj} and CMB-HD~\citep{CMB-HD:2022bsz} are likely to improve the prospects of probing helium reionization dramatically as suggested in Ref.~\citep{Hotinli:2022jna}.

Furthermore, the reionization of helium may potentially effect the selection functions of the high-redshift quasars and galaxies, as well as the star formation rate inferred from the CIB signal as the ionizing processes can modulate the ultra-violet background fluctuations and the absorption lines used for inferring redshifts with spectroscopic imaging surveys such as DESI and MegaMapper. Going forward, these effects should be accounted for and modelled for an unambiguous characterisation and detection of the helium reionization and can both act as increasing the high-redshift observables' sensitivity to helium reionization as well as potentially introducing biases or confusion.

The joint analysis of tracers of large-scale velocity fluctuations reconstructed from small-scale CMB and different LSS observables, large-scale density fluctuations from galaxy distributions, and reconstructed lensing potential may prove more challenging in practice than what we have considered here, as spurious correlations between these observables may arise in the case the same data (such as the same CMB maps) are used throughout. Also going forward, the prospects of jointly-analysing early structure formation with the methods highlighted here could be tested with realistic simulations including non-Gaussian foregrounds and systematics, and astrophysical properties including the choices made for the halo model of galaxies and electrons in this work could be taken into account consistently via a forward-modelling framework, for example. In what follows, we could make these advances to better identify the true prospects of characterising these epochs with cross-correlation studies.  

Finally, another difficulty posed by probing the epoch of helium reionization with the technique discussed here is that the mean electron fraction should also vary with the fraction of baryons that is locked up in astrophysical objects such as stars, stellar remnants (including baryons that
have disappeared into black holes), molecular and atomic clouds, and
any ionized systems that are optically thick to Thomson scattering. Such effects
are not taken into account in our modelling of the ionized fraction and may not be distinguished from measurements using SZ effects alone. The expected variations in the fraction of baryons turned to stars are of a few percent, comparable to the abundance of
helium by number, and the measurement of the mean electron fraction of the Universe is a combination of multitude of factors including reionization of hydrogen and helium, plus the fractional reservoirs of baryons in the HI damped Ly$\alpha$ absorption systems and stars, which are Thomson-thick repositories for ionized matter, constituting to a significant portion of baryons, and should be modeled together with helium reionization.

Nevertheless, our results are promising and should motivate further analysis of the prospects of cross-correlation science in the near future with ongoing stage-3 and upcoming stage-4 cosmology experiments. The epoch of helium reionization in particular carries valuable and novel information about astrophysics and cosmology that can potentially be accessed in the foreseeable future. In a series of upcoming works we will continue to explore the reach of this program, extending the observables and techniques introduced here.

\section{Acknowledgements}

We thank Fiona McCarthy, Matthew Johnson and Simone Ferraro for useful conversations. We thank Fiona McCarthy for her contributions to the conceptualisation of this project. SCH was supported by the Horizon Fellowship from Johns Hopkins University.

\bibliography{main}

\end{document}